\documentclass[10pt,journal,twocolumn]{IEEEtran}

\usepackage[dvips]{epsfig}
\usepackage[dvips]{graphicx}
\usepackage{amsmath,amsfonts,bm,amssymb,psfrag,ifthen,color,subfigure,algpseudocode,setspace,cite,stfloats,booktabs,float,url,indentfirst,tabularx,multirow}
\usepackage[ruled]{algorithm2e}
\usepackage{hyperref}

\newtheorem{lemma}{\bf Lemma}

\begin{document}
\title{Robust Power Allocation for UAV-aided ISAC Systems with Uncertain Location Sensing Errors}
\author{Junchang Sun, Shuai Ma, Ruixin Yang, Tingting Yang, and Shiyin Li
\thanks{Junchang Sun, Ruixin Yang, and Shiyin Li are with the School of Information and Control Engineering, China University of Mining and Technology, Xuzhou 221116, China (e-mail: sunjc@cumt.edu.cn, ray.young@cumt.edu.cn, lishiyin@cumt.edu.cn).}	
\thanks{Shuai Ma are with the Peng Cheng Laboratory, Shenzhen 518055, China (e-mail: mash01@pcl.ac.cn).}
\thanks{T. Yang is with the Peng Cheng Laboratory, Shenzhen 518000, China, and also with the School of Navigation, Dalian Maritime University, Dalian, 116026, China (e-mail: yangtt@pcl.ac.cn).}
}

\maketitle
\begin{abstract}
	Unmanned aerial vehicle (UAV) holds immense potential  in integrated sensing and communication (ISAC) systems for the Internet of Things (IoT). In this paper, we propose a UAV-aided ISAC framework and  investigate three robust power allocation schemes. 
	First, we derive an explicit expression of the  Cram\'er-Rao bound (CRB) based on time-of-arrival (ToA) estimation, which serves as the performance metric for location sensing. 
	Then, we analyze the impact  of the location sensing error (LSE)  on communications, revealing the inherent coupling relationship between communication and sensing. 
	Moreover, we formulate three robust communication and sensing power allocation problems by respectively characterizing the LSE as an ellipsoidal distributed model, a Gaussian distributed model, and an arbitrary distributed model. Notably, the optimization problems seek to minimize the CRB, subject to data rate and  total  power constraints. However, these problems are non-convex and intractable. To address the challenges related to the three aforementioned LSE models, we respectively propose to use the ${\cal{S}}$-Procedure and alternating optimization (${\cal{S}}$-AO) method, Bernstein-type inequality and successive convex approximation (BI-SCA) method, and  conditional value-at-risk (CVaR) and AO (CVaR-AO) method to solve these problems. 
	Finally, simulation results demonstrate the robustness of our proposed UAV-aided ISAC system against the LSE  by comparing with the non-robust design, and evaluate the trade-off between communication and sensing in the ISAC system.
\end{abstract}


\section{Introduction}
Integrated sensing and communication (ISAC) and unmanned aerial vehicle (UAV)  are emerged as key technologies for future six-generation (6G) networks \cite{Mu2023MCOM, Liu2022JSAC, Ma2023JIOT}. ISAC enables the communication and sensing signals to share the same frequency band, saving valuable  spectrum resources and effectively addressing the issue of spectrum scarcity. Moreover, integrating communication and sensing capabilities on a single device significantly reduces hardware costs \cite{Wei2023JIOT}. On the other hand, UAVs, due to their high mobility and  relatively high altitude, offer distinct advantages in establishing line-of-sight (LoS) aerial-to-ground (A2G) links, making them ideal for communication with users in challenging environments \cite{Mozaffari2019COMST}. Additionally,  leveraging millimeter wave and multi-antenna technologies, UAVs show great  potential in wireless sensing applications, particularly in emergency relief and military operations. 
The synergistic  advantages of both ISAC and UAV technologies motivate us to investigate UAV-aided ISAC systems, which hold vast potential in the Internet of Things (IoT).

The authors in \cite{Mu2023MCOM} provided a comprehensive overview of the challenges and future directions for UAV-aided ISAC systems. They notably emphasized the reciprocal benefits between communication and positioning.  Similarly, in \cite{Destino2017ICCW}, the authors  highlighted the mutually reinforcing nature of communication and sensing capabilities. 
Indeed, drawing from this insightful perspective, numerous studies have been undertaken to deeply explore the heightened performance potential  of UAV-aided ISAC systems. For instance, our previous work \cite{Sun2022LWC} optimized beamforming vectors by minimizing the transmit power for multi-input single-output (MISO) systems. This work illustrated that the integration of a priori sensing estimation information leads to improved communication performance. In \cite{Deng2023TWC}, the authors introduced an adaptable UAV-empowered ISAC mechanism. This innovation effectively improves the resource utilization rate and further enhances system throughput performance by jointly optimizing beamforming strategies and UAV trajectories. In \cite{Wang2021TCOMM}, a cooperative communication and sensing protocol was introduced, leveraging multi-UAV assistance. Here, the authors jointly optimized UAV locations, user associations, and power control to enhance network utility. Moreover, in \cite{Meng2022ICCW}, the authors embarked on a study that investigated a novel UAV-aided periodic ISAC system, offering a more flexible approach for analyzing communication and sensing dynamics.
However, the realization of an ISAC system brings with it the problem of mutual interference between communication and sensing \cite{Meng2023MWC, Wu2021JSAC}. Remarkably, existing studies have overlooked this critical aspect, thereby motivating  us to explore the inherent coupling relationship between communication and sensing results in an integrated framework.

Another critical consideration in the design of UAV-aided ISAC systems is to explore the mutual restrictions between communication and sensing in a power limited system. Consequently, numerous works have been carried out to analyze the trade-off between sensing and communication of ISAC systems. For instance, in our earlier work \cite{Sun2023TCOMM}, we introduced  a Ziv-Zakai bound-based ISAC system and improved its performance by optimizing an optimal power allocation problem. Moreover, we  conducted an in-depth analysis of the trade-off between communication and sensing. Similarly, in \cite{Fan2022LWC}, the authors  focused on power allocation optimization in a transmit power-limited ISAC network. In \cite{Dong2023TWC}, the authors performed a unified power allocation framework for the ISAC systems to flexibly allocate the limited power, and analyzed trade-off between communication and sensing services via various allocation schemes.
Moreover, to provide clarity regarding how ISAC systems effectively allocate power to simultaneously meet communication and sensing requisites, the authors in \cite{Ghatak2018VTCS} derived the trade-off function based on theoretical bounds. They meticulously designed the beamwidth and power splitting factor to cater to user requirements. Furthermore, within the domain of UAV-aided ISAC systems, \cite{Qin2023TWC} investigated a multi-UAV ISAC setup.  The authors devised a viable  power allocation policy to enhance system performance by maximizing the minimum weighted spectral efficiency. In a related vein, \cite{Liu2020WCNC} established power allocation strategy. This strategy minimizes the total transmit power while ensuring  detection performance and managing  latency violation probability in a UAV-enabled integrated radar and communication system.
It is imperative to acknowledge that the investigations mentioned above all rest on the assumption of accurate channel state information (CSI) availability for UAVs. However, the practical reality entails challenges in accurately obtaining CSI.  As highlighted in \cite{Garcia2019ACCESS}, the performance of ISAC systems is markedly sensitive to sensing outcomes. The precision of user sensing is hence pivotal for ensuring dependable communications. Regrettably, existing works tends to overlook the integration of sensing results into UAV communications within ISAC systems.   This drive serves as the motivation behind our pursuit of a robust design within a UAV-aided ISAC system. Our focus lies in systematically addressing the influence of the location sensing error (LSE) on communications.
 
To address the aforementioned challenges, in this paper, we reveal the fundamental relationship between the LSE and achievable communication rate. Building upon this understanding, we formulate optimization problems in terms of robust power allocation for UAV-aided ISAC systems. This endeavor involves the characterization of LSE using an ellipsoidal distributed model, a Gaussian distributed model, and an arbitrary distributed model. Accordingly, the main contributions of this work are summarized as follows:
\begin{itemize}
	\item[(1)] We propose a ISAC framework for UAV-aided systems that takes into account the uncertain LSE. First, we derive an explicit expression of the time-of-arrival (ToA)-based Cram\'er-Rao bound (CRB)  to evaluate the location sensing performance. Furthermore, we reveal  the inherent coupling relationship between communication and sensing by expressing the achievable rate as a function of the LSE. To the best of our knowledge, our work marks the inaugural revelation of the coupling between location sensing and the communication rate in UAV-aided ISAC systems.
	\item[(2)] To improve the performance of our proposed UAV-aided ISAC systems, we formulate three optimization problems designed for the robust communication and sensing power allocation. These problems aim to minimize the CRB subject to data rate and total transmit power constraints. Additionally, These problems are formulated by characterizing the LSE as different distribution models. The specific models employed for representing LSE are detailed as follows:
	\begin{itemize}
		\item[(i)] When the LSE is characterized as an ellipsoidal distributed model, we propose to use the ${\cal{S}}$-Procedure to conservatively transform the constraints into a finite number of linear matrix inequalities (LMIs). Subsequently, we apply the alternating optimization (AO) method to solve the optimization problem (namely, the ${\cal{S}}$-AO method).
		\item[(ii)] When the LSE is characterized as a Gaussian distributed model, the rate constraint becomes a probabilistic form, making the problem more intractable. In this case,	we propose to use the Bernstein-type inequality to conservatively transform this chance-constraint to a convex deterministic form. Subsequently, we apply the successive convex approximation (SCA)  method  to approximately solve the optimization problem (namely, the BI-SCA method).
		\item[(iii)] When the LSE is characterized as an arbitrary distributed model, similarly, the rate constraint is a probabilistic form. Even worse, we lack the a priori distribution information of the LSE, having access only to the first- and second-order moments. In this case, we propose to use the conditional value-at-risk (CVaR) method to transform the chance-constraint  into a deterministic form. Subsequently, we apply the AO method to solve the optimization problem iteratively (namely, the CVaR-AO method).
	\end{itemize}
	\item[(3)] Numerical simulations demonstrate the robust nature of the proposed UAV-aided ISAC system. We achieved this by conducting a thorough comparison with the non-robust design, showcasing the superior performance of our proposed approach. Furthermore, we reveal the trade-offs between communication and sensing by simulations. This investigation involved the variation of critical system parameters, allowing us to dissect the interplay between these essential aspects of the UAV-aided ISAC system.
\end{itemize}

The rest of the paper is organized as follows. In Section \ref{System_Model}, the system model, frame structure, signal transmission model, and the inherent couping relationship between the achievable rate and LSE are presented. In Section \ref{problem_formulation}, the problem formulation and corresponding algorithms are presented. In Section \ref{Simulation_Results}, numerical results are presented to demonstrate the performance of the proposed ISAC system. Finally, conclusions are provided in Section \ref{Conclusions}.

{\em Notations:} $a$, ${\bm{a}}$, ${\bm{A}}$, and ${\cal A}$ denote a scalar, vector, matrix, and set, respectively. $\Re \left\{  \cdot  \right\}$ denotes real part. $\left[ {\bm{a}} \right]_i$ denotes the $i$th element in the vector ${\bm{a}}$. The ${\rm{rank}}\left(  \cdot  \right)$, ${\rm{tr}}\left\{  \cdot  \right\}$, ${\left|  \cdot  \right|}$, ${\left\|  \cdot  \right\|}$, ${\left(  \cdot  \right)^{T}}$, ${\left(  \cdot  \right)^{H}}$, and ${\left(  \cdot  \right)^{-1}}$ denote rank, trace, absolute value, 2-norm, transpose, complex transpose, and inverse operations, respectively. 
${\bm{A}} \succeq {\bm{B}}$ means that matrix ${\bm{A}} - {\bm{B}}$ is positive semidefinite. ${\mathbb E}\left\{  \cdot  \right\}$ and $\Pr \left\{  \cdot  \right\}$ denote the expectation and the probability operator, respectively.  ${\bm{I}}$ is the identity matrix and ${\bm{1}}_{N}$ is a $N\times 1$ vector with all elements being ones. The other key notations and acronyms are listed in Table \ref{notations} and Table \ref{acronyms}, respectively.

\begin{table}[H]
	\centering
	\small
	\caption{Summary of Key Notations}
	\begin{tabular}{|l|l|}
		\hline
		Notations &  Description \\
		\hline
		${\hat{\bm{u}}}$ & Estimated location of the UE \\
		\hline
		$\Delta {\bm{u}}$ & LSE of the UE \\
		\hline
		${\hat{\bm{h}}}_k$ & Estimated CSI of the $k$th UAV\\
		\hline
		${\Delta {{\bm{h}}}_k}$ & CSI error of the $k$th UAV\\
		\hline
		$P_{{\rm{s}},k}$ & Sensing power in the $k$th UAV \\
		\hline
		$P_{{\rm{c}},k}$ & Communication power in the $k$th UAV \\
		\hline
		$P_{\rm{total}}$ & Total transmit power of UAVs \\
		\hline
		${{\bm{J}}_{\rm{p}}}$ & Location-related FIM \\
		\hline
		${R_k}$ & Achievable rate of the $k$th UAV\\
		\hline
		$P_{\rm{out}}$ & Outage probability \\
		\hline
		${\bar R}$ & Data rate requirement\\
		\hline
		$\delta$ & Ellipsoid parameter\\
		\hline
	\end{tabular}\label{notations}
\end{table}

\begin{table}[H]
	\centering
	\small
	\caption{Summary of Main Acronyms}
	\begin{tabular}{|l|l|}
		\hline
		Acronyms &  Description \\
		\hline 
		ISAC & Integrated sensing and communication\\
		\hline
		LSE & Location sensing error \\
		\hline
		UAV & Unmanned aerial vehicle \\
		\hline
		UE & User equipment \\
		\hline
		CRB & Cram\'er-Rao bound \\
		\hline
		AO & Alternating optimization \\
		\hline
		SCA & Successive convex approximation \\
		\hline
		CSI & Channel state information \\
		\hline
		FIM & Fisher information matrix \\
		\hline
		CVaR & Conditional value-at-risk\\
		\hline 
		LMIs & Linear matrix inequalities \\
		\hline
		SDP & Semidefinite program \\
		\hline
	\end{tabular}\label{acronyms}
\end{table}

\section{System Model}\label{System_Model}
\subsection{System Setup}
\begin{figure}[htbp]
	\centering
	\includegraphics[width=8cm]{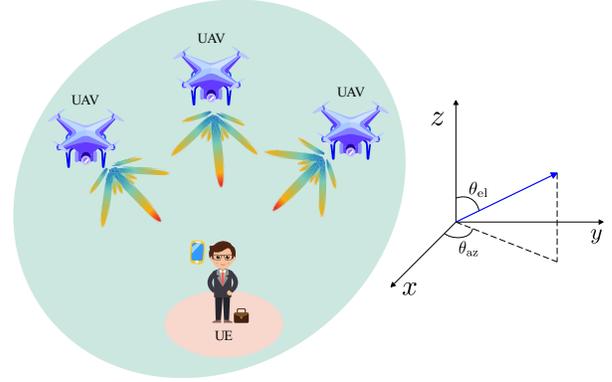}
	\caption{The illustration of the system model.}\label{sys_model}
\end{figure}

We consider a three-dimensional (3D) synchronous ISAC system containing $K$ multi-antenna UAVs and a single-antenna user equipment (UE), which is shown in Fig. \ref{sys_model}. Each UAV is equipped with  ${{N_t}}$ antenna elements of a uniform planar array (UPA). The known location of the $k$th UAV and the unknown location of the UE are denoted as ${{\bm{p}}_k} \in {\mathbb R}^3,k \in {{\cal K}}$ and ${\bm{u}} \in {\mathbb R}^3$, respectively, where ${{\cal K}} \buildrel \Delta \over = \left\{ {1, \cdots ,K} \right\}$. The location of the $m$th antenna element in the $k$th UAV is denoted as ${\bm{q}}_{k,m} \in {\mathbb R}^3$. The angle-of-departure (AoD) between the $k$th UAV and UE is denoted as ${{\bm \theta} _{k}} = \left[\theta_{{\rm{az}},k},\theta_{{\rm{el}},k}\right]^T$, where $\theta_{{\rm{az}},k}$ and $\theta_{{\rm{el}},k}$ are denoted as the AoD in azimuth and
elevation, respectively. According to the location geometric relationship, we have
\begin{subequations}\label{azel}
	\begin{align}
		\theta_{{\rm{az}},k} &\buildrel \Delta \over = {\rm{atan2}}\left( {{{\left[ {\bm{u}} \right]}_2} - {{\left[ {{{\bm{p}}_k}} \right]}_2},{{\left[ {\bm{u}} \right]}_2} - {{\left[ {{{\bm{p}}_k}} \right]}_1}} \right),\\
		{\theta _{{\rm{el}},k}} &\buildrel \Delta \over = {\rm{arccos}}\left( {\frac{{{{\left[ {\bm{u}} \right]}_3} - {{\left[ {{{\bm{p}}_k}} \right]}_3}}}{{\left\| {{\bm{u}} - {{\bm{p}}_k}} \right\|}}} \right).
	\end{align}
\end{subequations}

\subsection{Frame Structure}
\begin{figure}[htbp]
	\centering
	\includegraphics[width=8cm]{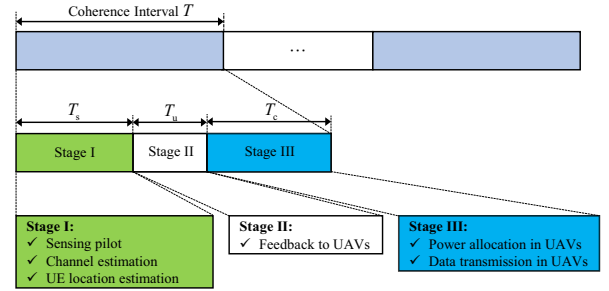}
	\caption{The illustration of the  frame structure of the ISAC system.} \label{signal_frame}
\end{figure}

As shown in Fig. \ref{signal_frame}, the system frame is divided into a downlink sensing pilot subframe (Stage I), an uplink feedback subframe (Stage II), and a downlink data transmission subframe (Stage III) \cite{Ma2022TCOMM}:
\begin{itemize}
 	\item At Stage I, i.e., the downlink sensing pilot period ${T_{\rm{s}}}$, UAVs transmit pilot signals to the UE. Then, the UE estimates the CSI and location sensing information{\footnote{We can estimate the location of the UE by using several unbiased positioning methods, such as the ML method \cite{Chan2006TVT, Cheung2004ICASSP} and the constrained weighted least-square method \cite{Cheung2004TSP}, which is not the main interest of this work.}}.
 	\item At Stage II, i.e., the uplink feedback period ${T_{\rm{u}}}$, the UE sends the feedback to UAVs, including the sensing information obtained during Stage I. 
 	\item At Stage III, i.e., the downlink data transmission period ${T_{\rm{c}}}$, UAVs design the power allocation strategy of communication and sensing based on the estimated CSI and UE location, and then transmit data signals to the UE. Note that the transmit signals from different UAVs are orthogonal to each other in the frequency domain.
\end{itemize}

\subsection{Sensing Model}
In the sensing period $T_{\rm{s}}$, the transmitted sensing pilot signal ${{\bm{x}}_{{\rm{s}},k}}\left( t \right)$ from the $k$th UAV is modeled as
\begin{align}
{{\bm{x}}_{{\rm{s}},k}}\left( t \right) = {\sqrt{P_{{\rm{s}},k}}} {{\bm{w}}_k}{s_{{\rm{s}},k}}\left( t \right),
\end{align}
where ${P_{{\rm{s}},k}}$ and ${{\bm{w}}_k}$ denote the transmitted sensing power and beamforming vector, respectively, where $\left\| {\bm{w}}_k \right\|^2 = 1$, and ${s_{{\rm{s}},k}}\left( t \right)$ denotes the pilot symbol with the unit amplitude.

Due to the fact that UAVs operate at a relatively high altitude, there are typically LoS A2G connection links between the UAVs and ground UE, which is widely adopted by previous works \cite{Khuwaja2018COMST, Wu2018JSAC}. In this case, we only consider the LoS condition and denote the received sensing signal ${y_{{\rm{s}},k}}\left( t \right) $  from the $k$th UAV as
\begin{align}\label{receive_pos_signal}
{y_{{\rm{s}},k}}\left( t \right) ={\sqrt{P_{{\rm{s}},k}}} {\bm{h}}_k^H{{\bm{w}}_k}{s_{{\rm{s}},k}}\left( {t - {\tau _k}} \right) + {n}\left( t \right),
\end{align}
where ${\tau _k} = \frac{{\left\| {{{\bm{u}}} - {{\bm{p}}_k}} \right\|}}{c}$ represents the time delay between the $k$th UAV and UE for the synchronous system, $c$ denotes the speed of light, ${n}\left( t \right)$ is the additive white Gaussian noise (AWGN) with zero-mean and two-sided power spectral density ${N_0}/2$.

Moreover, the complex channel  ${{\bm{h}}_k}$ between the $k$th UAV and UE is given by
\begin{align}
{{\bm{h}}_k} = {\rho _k}{\bm{a}}\left( {{\bm{\theta} _k}} \right),
\end{align}
where ${\rho _k} = {\left| {\rho _k} \right|}{e^{j\psi_k}}$ denotes the complex channel gain, ${\psi_k}$ denotes the phase shift, and ${\bm{a}}\left( {{\bm{\theta} _k}} \right)$ is the antenna response vector, given by
\begin{align}
	{\bm{a}}\left( {\bm{\theta}_k} \right) = {e^{j{{\bm{Q}}_k^T}{\bm{\kappa }}\left( {{{\bm{\theta}} _k}} \right)}},
\end{align}
where ${{\bm{Q}}_k} \buildrel \Delta \over = \left[ {{{\bm{q}}_{k,1}}, \cdots ,{{\bm{q}}_{k,{N_t}}}} \right]$, and
\begin{align}
	{{\bm{\kappa }}\left( {{{\bm{\theta}} }} \right)} \buildrel \Delta \over = \frac{{2\pi }}{\lambda }{\left[ {\sin {\theta _{{\rm{el}}}}\cos {\theta _{{\rm{az}}}},\sin {\theta _{{\rm{el}}}}\sin {\theta _{{\rm{az}}}},\cos {\theta _{{\rm{el}}}}} \right]^T},
\end{align}
where $\lambda$ denotes the wave length. For the notation convenience, we denote by ${{\bm{a}}_k} \buildrel \Delta \over = {\bm{a}}\left( {{\bm{\theta} _k}} \right)$.

In this work, we propose to use the CRB of the LSE to evaluate the sensing performance in the ISAC system. The CRB is basically used as the theoretical lower bound for unbiased estimators. 
Hence, based on the received sensing signal ${y_{{\rm{s}},k}}\left( t \right) $, we calculate the location-related CRB as
\begin{align}
{\rm{CRB}} = {\rm{tr}}\left\{ {{\bm{J}}_{\rm{p}}^{-1}}\right\},
\end{align}
where ${\bm{J}}_{\rm{p}}$ is the location-related Fisher information matrix (FIM), expressed as
\begin{align}\label{location_related_fim}
{{\bm{J}}_{\rm{p}}} = \frac{{8{\pi ^2}{\beta}^2}}{{{c^2}{N_0}}}\sum\limits_{k \in {{\cal K}}} {{P_{{\rm{s}},k}{\left| {{\bm{h}}_k^H}{{\bm{w}}_k} \right|}^2}{\bm{\Psi }}\left( {{\bm{\theta} _k}} \right)},
\end{align}
where $\beta$ is the effective bandwidth and 
\begin{small}
\begin{align}
&{\bm{\Psi }}\left( {\bm{\theta}}  \right) \buildrel \Delta \over  = \nonumber\\
&\left[ {\begin{array}{*{20}{c}}
	{{{\left( {\sin {\theta _{{\rm{el}}}}\cos {\theta _{{\rm{az}}}}} \right)}^{^2}}}&{\frac{1}{2}{{\sin }^2}{\theta _{{\rm{el}}}}\sin 2{\theta _{{\rm{az}}}}}&{\frac{1}{2}\sin 2{\theta _{{\rm{el}}}}\cos {\theta _{{\rm{az}}}}}\\
	{\frac{1}{2}{{\sin }^2}{\theta _{{\rm{el}}}}\sin 2{\theta _{{\rm{az}}}}}&{{{\left( {\sin {\theta _{{\rm{el}}}}\sin {\theta _{{\rm{az}}}}} \right)}^{^2}}}&{\frac{1}{2}\sin 2{\theta _{{\rm{el}}}}\sin {\theta _{{\rm{az}}}}}\\
	{\frac{1}{2}\sin 2{\theta _{{\rm{el}}}}\cos {\theta _{{\rm{az}}}}}&{\frac{1}{2}\sin 2{\theta _{{\rm{el}}}}\sin {\theta _{{\rm{az}}}}}&{{{\cos }^2}{\theta _{{\rm{el}}}}}
	\end{array}} \right].
\end{align}
\end{small}%
The details of the derivation of ${\bm{J}}_{\rm{p}}$ can be found in Appendix.

\subsection{Communication Model}
In the communication period ${T_{\rm{c}}}$, the transmitted data transmission signal ${{\bm{x}}_{{\rm{c}},k}}\left( t \right)$ from the $k$th UAV is modeled by 
\begin{align}
{{\bm{x}}_{{\rm{c}},k}}\left( t \right) ={\sqrt{P_{{\rm{c}},k}}} {{\bm{w}}_k}{s_{{\rm{c}},k}}\left( t \right),
\end{align}
where ${P_{{\rm{c}},k}}$ denotes the transmitted communication power and ${s_{{\rm{c}},k}}\left( t \right)$ denotes the data symbol, which satisfies normalized energy. It is worth pointing out that we use the same beamforming vector for the data transmission as for the sensing.

The estimated CSI ${{{{\bm{\hat h}}_k}}}$ and location ${\bm{\hat u}}$ of the UE are available in UAVs after the downlink sensing pilot period ${T_{\rm{s}}}$ and the uplink feedback period ${T_{\rm{u}}}$. Subsequently, based on the path loss model of $\left|{\rho }_k \right| = \frac{\lambda }{{4\pi {{\left\| {{\bm{ u}} - {{\bm{p}}_k}} \right\|}}}}$ \cite{Abu_Shaban2018TWC, Sun2022LWC}, we can write  ${{{{\bm{\hat h}}_k}}}$ as
\begin{align}
{{\bm{\hat h}}_k} = \frac{{\lambda {e^{j\psi_k}} }}{{4\pi }}\frac{1}{{{{\left\| {{\bm{\hat u}} - {{\bm{p}}_k}} \right\|}}}}{\bm{\hat a}}_k, \label{channel}
\end{align}
where ${\bm{\hat a}}_k$ is the estimated antenna array response vector, which can be constructed based on the estimated location ${\bm{\hat u}}$ by \eqref{azel}.

The received data transmission signal ${y_{{\rm{c}},k}}\left( t \right)$ from the $k$th UAV is given by
\begin{align}
{y_{{\rm{c}},k}}\left( t \right) = {\sqrt{P_{{\rm{c}},k}}}{\left( {{{{\bm{\hat h}}}_k} + \Delta {{\bm{h}}_k}} \right)^H}{{\bm{w}}_k}{s_{{\rm{c}},k}}\left( {t - {\tau _k}} \right) + n\left( t \right),
\end{align}
where ${\Delta {{\bm{h}}}_k}$ represents the CSI error, which can be  mathematically written as
\begin{align}
\Delta {{\bm{h}}}_k = \frac{{\lambda {e^{j\psi_k}}{\bm{\hat a}}_k }}{{4\pi }}\left( {\frac{1}{{{{\left\| {{\bm{\hat u}} + \Delta {\bm{u}} - {{\bm{p}}_k}} \right\|}}}} - \frac{1}{{{{\left\| {{\bm{\hat u}} - {{\bm{p}}_k}} \right\|}}}}} \right),\label{channe2}
\end{align}
where $\Delta {\bm{u}} \buildrel \Delta \over = {\bm{u}} - {\bm{\hat u}}$ is denoted as the LSE.

When the accurate location of the UE is unknown, we can mathematically calculate the estimated achievable rate, which is influenced by the allocated communication power and the LSE. Based on \eqref{channel} and \eqref{channe2}, according to Shannon's theorem, the estimated achievable rate ${R_k}$ (bps/Hz)  of the $k$th UAV in the downlink data transmission period is given by
\begin{align}
{R_k} &= {\log _2}\left( {1 + \frac{{{P_{{\rm{c}},k}}{{\left| {{{\left( {{{{\bm{\hat h}}}_k} + \Delta {{\bm{h}}_k}} \right)}^H}{{\bm{w}}_k}} \right|}^2}}}{{{N_0}}}} \right) \\
&= {\log _2}\left( {1 + \frac{{{P_{{\rm{c}},k}}{\lambda ^2}{{\left| {{\bm{\hat a}}_k^H{{\bm{w}}_k}} \right|}^2}}}{{16{\pi ^2}{N_{\rm{0}}}{{\left\| {{\bm{\hat u}} + \Delta {\bm{u}} - {{\bm{p}}_k}} \right\|}^2}}}} \right).\label{rate}
\end{align}

\section{Robust Power Allocation for the ISAC Design}\label{problem_formulation}
In practice, the accurate CSI and location sensing information are difficult to be obtained. Hence, considering the impact of the LSE, it is reasonable  to perform robust designs within the tolerance of uncertainty.
In this section, we aim to design three robust communication and sensing power allocation schemes to improve the performance of our proposed UAV-aided ISAC framework. Specifically, we minimize the CRB subject to the achievable rate and total power constraints with three types of LSE distribution models, namely, an ellipsoidal distributed model, a Gaussian distributed model, and  an arbitrary distributed model. Additionally, we provide a non-robust ISAC design scheme to serve as a basis for comparisons.

\subsection{Robust ISAC Design with the Ellipsoidal Distributed LSE}
In this subsection, we first consider that the LSE $\Delta {\bm{u}}$ is bounded. To be more specific, the ellipsoidal model is commonly used to quantify a bounded error distribution model \cite{Pascual2006TSP, Ma2013LCOMM}, which makes the formulated optimization problem tractable. Hence, we assume that the LSE $\Delta {\bm{u}}$ falls within a FIM-related ellipsoidal region ${\cal{R}}$, expressed as \cite{Torrieri1984TAES}
\begin{align}
{\cal R} \buildrel \Delta \over = \left\{ {\Delta {\bm{u}}|\Delta {{\bm{u}}^T}{\bm{J}}_{\rm{p}}\Delta {\bm{u}} \le \delta} \right\},
\end{align}
where $\delta$ is the ellipsoidal parameter, which determines the volume of the ellipsoid.

In this case, the robust power allocation problem with the  ellipsoidal distributed LSE is formulated as
\begin{subequations}\label{Bound}
	\begin{align}
	\mathop {\min }\limits_{P_{{\rm{s}},k} ,P_{{\rm{c}},k} }&\quad {\rm{tr}}\left\{ {{\bm{J}}_{\rm{p}}^{ - 1}} \right\} \label{Bound_1} \\
	{\rm{s.t.}}&\quad {{R_k} \ge \bar R},k \in {{\cal K}}, \label{Bound_2} \\
	&\quad \Delta {{\bm{u}}^T}{\bm{J}}_{\rm{p}}\Delta {\bm{u}} \le \delta, \label{Bound_3} \\
	&\quad \sum\limits_{k \in {{\cal{K}}}} {{P_{{\rm{s}},k}} + {P_{{\rm{c}},k}}} \le P_{\rm{total}},\label{Bound_4}\\
	&\quad {{\bm{J}}}_{\rm{p}} \succeq {\bm{0}}, \label{Bound_5}
	\end{align}
\end{subequations}
where ${\bar R}$ denotes the data rate requirement and $P_{\rm{total}}$ is the total power carried by UAVs. However, the optimization problem is non-convex and intractable. In the following, we propose to use the ${\cal{S}}$-Procedure and AO (${\cal{S}}$-AO) method to solve this problem. 

Specifically, by substituting \eqref{rate} into \eqref{Bound_2}, we transform  problem \eqref{Bound} into 
\begin{subequations}\label{BoundSDR}
	\begin{align}
	\mathop {\min }\limits_{P_{{\rm{s}},k},P_{{\rm{c}},k}} &\quad {\rm{tr}}\left\{ {{\bm{J}}_{\rm{p}}^{ - 1}} \right\} \label{BoundSDR_1} \\
	{\rm{s}}.{\rm{t}}.&\quad  {\left\| {\Delta {\bm{u}}} \right\|^2} + 2\Delta {{\bm{u}}^T}\left( {{\bm{\hat u}} - {{\bm{p}}_k}} \right) \le  {{\gamma_k}{P_{{\rm{c}},k}}} \nonumber \\ 
	&\qquad\qquad\qquad\qquad  - \left\| {{\bm{\hat u}} - {{\bm{p}}_k}} \right\|^2,k \in {{\cal K}}, \label{BoundSDR_2} \\
	&\quad \eqref{Bound_3}, \eqref{Bound_4}, \eqref{Bound_5}, \nonumber
	\end{align}
\end{subequations}
where ${\gamma_k } \buildrel \Delta \over = \frac{{\lambda ^{\rm{2}}}\left|{\bm{\hat a}}_k^H{{\bm{w}}_k}\right|^2}{ {16{\pi ^{\rm{2}}}{N_0}\left( {{2^{{{\bar R}}}} - 1} \right)}}$. It is easy to find that  problem \eqref{BoundSDR} is still non-convex and hard to be solved due to the constraints \eqref{Bound_3} and \eqref{BoundSDR_2}.  Hence, we adopt the following lemma to conservatively transform the constraints \eqref{Bound_3} and \eqref{BoundSDR_2} into a finite number of LMIs.

\begin{lemma}{(${\cal{S}}$-Procedure)} \cite{Boyd2004} \label{s_Procedure} 
	Let ${f_i}\left( {\bm{x}} \right) = {{\bm{x}}^T}{{\bm{A}}_i}{\bm{x}} + 2\Re \left\{ {{{\bm{x}}^T}{{\bm{b}}_i}} \right\} + {c_i}$ for $i \in \left\{ {0,1} \right\}$, where ${{\bm{A}}_i} \in {{\mathbb{H}}^{N \times N}}$, ${{\bm{x}}}\in {{\mathbb{C}}^{N}}$, ${{\bm{b}}_i} \in {{\mathbb{C}}^{N}}$, and ${c}_i \in {{\mathbb{R}}}$. Then, if there exists a vector ${\hat{\bm x}}$ such that ${f_1}\left( {{\bm{\hat x}}} \right) < 0$ holds, the following statements are equivalent:
	
	1. ${f_0}\left( {{\bm{ x}}} \right) \ge 0$ for all ${\bm{x}}$ such that ${f_1}\left( {{\bm{ x}}} \right) \le 0$.
	
	2. There exists a $\lambda_s \ge 0$ such that
	\begin{align}
	\left[ {\begin{array}{*{20}{c}}
		{{{\bm{A}}_0}}&{{{\bm{b}}_0}}\\
		{{\bm{b}}_0^T}&{{c_0}}
		\end{array}} \right] + {\lambda _s}\left[ {\begin{array}{*{20}{c}}
		{{{\bm{A}}_1}}&{{{\bm{b}}_1}}\\
		{{\bm{b}}_1^T}&{{c_1}}
		\end{array}} \right] \succeq {\bm{0}}.
	\end{align}
\end{lemma}

Based on the ${\cal{S}}$-Procedure in Lemma \ref{s_Procedure},  problem \eqref{BoundSDR} is reformulated as 
\begin{subequations}\label{sub}
	\begin{align}
	\mathop {\min }\limits_{ {P_{{\rm{s}},k},P_{{\rm{c}},k},{\lambda_s}}} &\quad  {\rm{tr}}\left\{ {{\bm{J}}_{\rm{p}}^{ - 1}} \right\} \label{s1_1} \\
	{\rm{s}}.{\rm{t}}.&\quad {\lambda _s}{{\bm{A}}_{1}} -{{\bm{A}}_{0,k}} \succeq {\bm{0}},k \in {{\cal K}}, \label{s1_2} \\
	&\quad \eqref{Bound_4}, \eqref{Bound_5}. \nonumber
	\end{align}
\end{subequations}
where 
\begin{align}
{{\bm{A}}_{0,k}} &= \left[ {\begin{array}{*{20}{c}}
	{{\bm{I}}}&{{\bm{\hat u}} - {{\bm{p}}_k}}\\
	{{{\left( {{\bm{\hat u}} - {{\bm{p}}_k}} \right)}^T}}&{-{{\gamma_k}{P_{{\rm{c}},k}}} + \left\| {{\bm{\hat u}} - {{\bm{p}}_k}} \right\|^2}
	\end{array}} \right],\\
{{\bm{A}}_{1}} &= \left[ {\begin{array}{*{20}{c}}
	{{\bm{J}}_{\rm{p}}}&{\bm{0}}\\
	{\bm{0}}^T&-{\delta}
	\end{array}} \right].
\end{align} 

Due to the fact that the variable $\lambda_s$ is coupled with the variable $P_{{\rm{s}},k}$ in ${\bm{A}}_1$, it is difficult to solve  problem \eqref{sub} directly. However, this problem can be decomposed into two convex subproblems with two variables $P_{{\rm{s}},k}$ and $P_{{\rm{c}},k}$. For the fixed $P_{{\rm{s}},k}$ (or, $P_{{\rm{c}},k}$), the corresponding subproblem can be transformed into convex forms. Therefore, we subsequently propose to adopt the AO algorithm to solve  problem \eqref{BoundSDR}, which can guarantee to
globally converge to the stationary point \cite{Grippof1999globally, Grippo2000}. Specifically, at the $n$th iteration, the two subproblems are optimized as follows:

\subsubsection{Communication Subproblem} For the given sensing power $P_{{\rm{s}},k}^{\left(n-1\right)}$, the communication power $P_{{\rm{c}},k}^{\left(n\right)}$ and variable $\lambda_s^{\left(n\right)}$ can be updated by solving
\begin{subequations}\label{subp1}
	\begin{align}
	\mathop {\min }\limits_{P_{{\rm{c}},k},{\lambda_s}} &\quad \sum\limits_{k \in {{\cal{K}}}} {{P_{{\rm{c}},k}}} \label{subp1_1} \\
	{\rm{s}}.{\rm{t}}.&\quad \eqref{s1_2}.\nonumber
	\end{align}
\end{subequations}

\subsubsection{Sensing Subproblem} For the obtained communication power $P_{{\rm{c}},k}^{\left(n\right)}$ and parameter $\lambda_s^{\left(n\right)}$, the sensing power $P_{{\rm{s}},k}^{\left(n\right)}$ can be updated by solving
\begin{subequations}\label{subp2}
	\begin{align}
	\mathop {\min }\limits_{ P_{{\rm{s}},k}} &\quad  {\rm{tr}}\left\{ {{\bm{J}}_{\rm{p}}^{ - 1}}\right\}  \label{subp2_1} \\
	{\rm{s}}.{\rm{t}}.&\quad \eqref{Bound_4}, \eqref{Bound_5}, \eqref{s1_2}. \nonumber
	\end{align}
\end{subequations}
We can adopt an interior point method to solve the above two subproblems \eqref{subp1} and \eqref{subp2}, based on off-the-shelf convex optimization solvers such as CVX \cite{Grant2014}. 
The details are listed in Algorithm \ref{alg1}.
\begin{algorithm}[htp]
	\caption{${\cal{S}}$-AO method for the robust power allocation with the ellipsoidal distributed LSE}
	\label{alg1}
	\LinesNumbered
	\SetKwRepeat{Repeat}{repeat}{until}
	\SetKw{Ini}{Initialization:}
	\KwIn{$\bar R > 0$, $\delta > 0$, $P_{\rm{total}} > 0$, the tolerance of accuracy $\epsilon > 0$, and the maximum iteration number $N_{\rm{max}}$}
	\KwOut{Sensing power $P_{{\rm{s}},k}$ and communciation power $P_{{\rm{c}},k}$}
	\Ini $P_{{\rm{s}},k}^{\left(0\right)} > 0$ and set $n=0$\;
	\Repeat{The decrease of the objective value is below the given tolerance $\epsilon$ {\rm{or}} $n > N_{\rm{max}}$}{
		$n = n + 1$\;
		Update $\left\{P_{{\rm{c}},k}^{\left(n \right)}, \lambda_s^{\left(n\right)}\right\}$ by solving  subproblem \eqref{subp1}\;
		Update $P_{{\rm{s}},k}^{\left(n \right)}$ by solving  subproblem \eqref{subp2}\;
	}
	\KwRet $P_{{\rm{s}},k}$ and $P_{{\rm{c}},k}$\;
\end{algorithm}

\subsection{Robust ISAC Design with the Gaussian Distributed LSE}
In this subsection, we characterize the LSE ${\Delta {\bm{u}}}$ as a Gaussian distribution model with zero-mean and  variance ${\bm{J }}_{\rm{p}}^{ - 1}$ \cite{Wang2009TSP, Lottici2002, Gezici2005MSP}.	The assumption of this distribution model is attractive due to the fact that the Gaussian distribution has outstanding mathematical properties, and thus simplifies the theoretical analysis. Moreover, the Gaussian distribution assumption is reasonable because the CRB can be achieved by several unbiased estimators.

In this case, the robust power allocation problem with the  Gaussian distributed LSE  is formulated as
\begin{subequations}\label{p1Gau}
	\begin{align}
	\mathop {\min }\limits_{P_{{\rm{s}},k} ,P_{{\rm{c}},k} }&\quad {\rm{tr}}\left\{ {{\bm{J}}_{\rm{p}}^{ - 1}} \right\} \label{p1Gau_1} \\
	{\rm{s}}.{\rm{t}}.&\quad \Pr \left\{ {{R_k}\le \bar R} \right\} \le {P_{\rm out}},k \in {{\cal K}}, \label{p1Gau_2} \\
	&\quad \Delta {\bm{u}} \sim {\cal N}\left( {{\bm{0}},{\bm{J}}_{\rm{p}}^{ - 1}} \right), \label{p1Gau_3}\\
	&\quad \sum\limits_{k \in {{\cal{K}}}} {{P_{{\rm{s}},k}} + {P_{{\rm{c}},k}}} \le P_{\rm{total}},\label{p1Gau_4}\\
	&\quad {{\bm{J}}}_{\rm{p}} \succeq {\bm{0}}, \label{p1Gau_5}
	\end{align}
\end{subequations}
where $P_{\rm{out}}$ denotes the tolerable outage probability. It can be observed that  problem \eqref{p1Gau} is intractable to be solved due to the chance-constraint \eqref{p1Gau_2} and the constraint \eqref{p1Gau_3}.  In the following, we propose to use the Bernstein-type inequality and SCA (BI-SCA) method to solve this problem.

First, we propose the following lemma to reformulate this non-convex problem.
\begin{lemma}{(Bernstein-type inequality)} \cite{Wang2014, Bechar2009}\label{lemma2}
	 Let $f = {{\bm{x}}^T}{\bm{Ax}} + 2\Re \left\{ {{{\bm{x}}^T}{\bm{a}}} \right\}$, where ${\bm{A}} \in {{\mathbb H}^N}$, ${{\bm{a}} \in {{\mathbb R}^N}}$, and ${\bm{x}} \sim {\cal N}\left( {{\bm{0}},{\bm{I}}} \right)$. Next, for a given constant $\zeta  > 0$, we can obtain
\begin{small}
	\begin{align}
	\Pr \left\{ {f  \ge {\rm{tr}}\left\{ {\bm{A}} \right\} + \sqrt {2\zeta } \sqrt {\left\| {\bm{A}} \right\|_{\rm{F}}^2 + 2\left\| {\bm{a}} \right\|_2^2}  + \zeta {{\lambda}^ + }\left( {\bm{A}} \right)} \right\} \le {e^{ - \zeta }},
	\end{align}
\end{small}%
where ${{\lambda ^ + }\left( {\bm{A}} \right) = \max \left\{ {{\lambda _{\max }}\left( {\bm{A}} \right),0} \right\}}$ and ${{\lambda _{\max }}\left( {\bm{A}} \right)}$ is the maximum eigenvalue of the matrix ${\bm{A}}$.
\end{lemma}

Based on the Lemma \ref{lemma2}, by substituting \eqref{rate} into \eqref{p1Gau_2}, the constraints \eqref{p1Gau_2} and \eqref{p1Gau_3} are jointly changed into
\begin{subequations}\label{Bernstein_express}
	\begin{align}
	&{\rm{tr}}\left\{ {{\bm{J}}_{\rm{p}}^{ - 1}} \right\} + \sqrt {2\eta } \omega_k  + \eta \varrho  - {{\gamma _k}{P_{{\rm{c}},k}}} + \left\| {{\bm{\hat u}} - {{\bm{p}}_k}} \right\|^2  \le 0, \label{constraint_b_1} \\ 
	&{\left\| {\begin{array}{*{20}{c}}
	{{\rm{vec}}\left( {{{\bm{J}}_{\rm{p}}^{ - 1}}} \right)}\\
	{\sqrt 2 {{\bm{J}}_{\rm{p}}^{ - \frac{1}{2}}}\left({{\bm{\hat u}} - {{\bm{p}}_k}}\right)}\end{array}} \right\|^2} \le \omega _k^2, \label{constraint_b_2}\\
	&\varrho {\bm{I}} - {{\bm{J}}_{\rm{p}}^{ - 1}} \succeq {\bm{0}}, \label{constraint_b_3}
	\end{align}
\end{subequations}
where $\eta  =  - \ln \left( {{P_{\rm out}}} \right)$, $\omega_k > 0$ and $\varrho > 0$ are slack variables.

It can be observed that the constraints \eqref{constraint_b_2} and \eqref{constraint_b_3} are still non-convex with respect to the sensing power $P_{{\rm{s}},k}$. Subsequently, we use the SCA method to solve problem \eqref{p1Gau} approximately. For the given initial values $\left\{ P_{{\rm{s}},k}^{\left( 0 \right) },{\omega _k^{\left( 0 \right)}} \right\}$, the iteration of the optimization problem is expressed as 
\begin{subequations}\label{p1Gau_solve}
	\begin{align}
	\mathop {\min }\limits_{P_{{\rm{s}},k} ,P_{{\rm{c}},k},\omega_k ,\varrho }&\quad {\rm{tr}}\left\{ {{\bm{J}}_{\rm{p}}^{ - 1}} \right\} \label{p1Gau_solve1} \\
	{\rm{s.t.}}\quad&{f_1}\left( P_{{\rm{s}},k} \right) + 2{\left( {{\bm{\hat u}} - {{\bm{p}}_k}} \right)^T}{{\bm{\Omega }}}\left( P_{{\rm{s}},k} \right)\left( {{\bm{\hat u}} - {{\bm{p}}_k}} \right) \nonumber\\
	&\qquad\qquad\qquad - {f_2}\left( {{\omega _k}} \right) \le 0, k \in {{\cal K}},\label{p1Gau_solve2}\\
	&{\varrho} {\bm{I}} - {{\bm{\Omega}}}\left(  P_{{\rm{s}},k}\right) \succeq {\bm{0}}, \label{p1Gau_solve3} \\
	&\eqref{p1Gau_4}, \eqref{p1Gau_5}, \eqref{constraint_b_1}, \nonumber
	\end{align}
\end{subequations}
where
\begin{subequations}
	\begin{align}
	{f_1}\left( P_{{\rm{s}},k} \right) &\buildrel \Delta \over = {\rm{tr}}\left\{ {{\bm{J}}_{{\rm{p}},0}^{ - 2}} \right\}
	-2{\rm{tr}}\left\{ {{\bm{J}}_{{\rm{p}},0}^{ - 3}\left( {{{\bm{J}}_{\rm{p}}} - {{\bm{J }}_{{\rm{p}},0}}} \right)} \right\},\\
	{{\bm{\Omega }}}\left( P_{{\rm{s}},k} \right) &\buildrel \Delta \over = {\bm{J}}_{{\rm{p}},0}^{ - 1} - {\bm{J}}_{{\rm{p}},0}^{- 2} \left( {{{\bm{J}}_{\rm{p}}} - {\bm{J}}_{{\rm{p}},0}} \right),\\
	{f _2}\left( {{\omega _k}} \right) &\buildrel \Delta \over = {\left( \omega _k^{\left(0\right)} \right)^2} + 2\omega _k^{\left(0\right)}\left( {\omega _k} - \omega _k^{\left(0\right)} \right),
	\end{align}
\end{subequations}
where ${\bm{J}}_{{\rm{p}},0}$ is calculated based on the given $P_{{\rm{s}},k}^{\left(0\right)}$. Finally, this problem can be directly solved by CVX. The details are listed in Algorithm \ref{alg0}.
\begin{algorithm}[htp]
	\caption{BI-SCA method for the robust power allocation  with the Gaussian distributed LSE}
	\label{alg0}
	\LinesNumbered
	\SetKwRepeat{Repeat}{repeat}{until}
	\SetKw{Ini}{Initialization:}
	\KwIn{$\bar R > 0$, ${P_{\rm{out}}} > 0$,  $P_{\rm{total}} > 0$, the tolerance of accuracy $\epsilon > 0$, and the maximum iteration number $N_{\rm{max}}$}
	\KwOut{Sensing power $P_{{\rm{s}},k}$ and communciation power $P_{{\rm{c}},k}$}
	\Ini $P_{{\rm{s}},k}^{\left(0\right)} > 0$ and set $n=0$\;
	\Repeat{The decrease of the objective value is below the given tolerance $\epsilon$ {\rm{or}} $n > N_{\rm{max}}$}{
		$n = n + 1$\;
		Calculate ${\bm{J}}_{{\rm{p}},n-1}$ based on  $P_{{\rm{s}},k}^{\left( n-1 \right)}$\;
		Obtain  $P_{{\rm{s}},k}^{\left(n \right)}$,  $P_{{\rm{c}},k}^{\left(n \right)}$, and ${\omega_k ^{\left( n \right)}}$ as solutions of  problem \eqref{p1Gau_solve}\;
	}
	\KwRet  $P_{{\rm{s}},k}$ and $P_{{\rm{c}},k}$\;
\end{algorithm}

\subsection{Robust ISAC Design with the Arbitrary Distributed LSE}\label{Arbitrary_problem}
More practically, the distribution of the LSE $\Delta {\bm{u}}$ is unknown, i.e., the distribution of the LSE $\Delta {\bm{u}}$ is arbitrary. In contrast, it is easier to obtain the first- and second-order moments of $\Delta {\bm{u}}$. As mentioned above, the CRB is achievable. Thus, we can obtain that ${\mathbb{E}}\left\{ {\Delta {\bm{u}}} \right\} = {\bm{0}}$ and ${\mathbb{V}}\left\{ {\Delta {\bm{u}}} \right\} = {\bm{J}}_{\rm{p}}^{ - 1}$, where ${\mathbb{V}}\left\{ \cdot \right\}$ represents the variance operator.

In this case, the robust power allocation problem with the  arbitrary distributed LSE  is formulated as
\begin{subequations}\label{p1Ar}
	\begin{align}
	\mathop {\min }\limits_{P_{{\rm{s}},k} ,P_{{\rm{c}},k} }&\quad {\rm{tr}}\left\{ {{\bm{J}}_{\rm{p}}^{ - 1}} \right\} \label{p1Ar_1} \\
	{\rm{s}}.{\rm{t}}.&\quad \Pr \left\{ {{R_k}\le \bar R} \right\} \le {P_{\rm out}},k \in {{\cal K}}, \label{p1Ar_2} \\
	&\quad {\mathbb{E}}\left\{ {\Delta {\bm{u}}} \right\} = {\bm{0}}, \label{p1Ar_3}\\
	&\quad {\mathbb{V}}\left\{ {\Delta {\bm{u}}} \right\} = {\bm{J}}_{\rm{p}}^{ - 1}, \label{p1Ar_4}\\
	&\quad \sum\limits_{k \in {{\cal{K}}}} {{P_{{\rm{s}},k}} + {P_{{\rm{c}},k}}} \le P_{\rm{total}},\label{p1Ar_5}\\
	&\quad {{\bm{J}}}_{\rm{p}} \succeq {\bm{0}}. \label{p1Ar_6}
	\end{align}
\end{subequations}
In the following, we propose to use the CVaR and AO (CVaR-AO) method  to solve this intractable optimization  problem.

First, we substitute \eqref{rate} into the constraint \eqref{p1Ar_2}, given by
\begin{align}\label{Ar_exp1}
	\Pr \left\{ {\Delta {{\bm{u}}^T}\Delta {\bm{u}} + 2\Delta {{\bm{u}}^T}\left( {{\bm{\hat u}} - {{\bm{p}}_k}} \right) + {{\left\| {{\bm{\hat u}} - {{\bm{p}}_k}} \right\|}^2} - {P_{{\rm{c}},k}}{\gamma _k} \le 0} \right\}\nonumber \\
	 \ge 1 - {P_{{\rm{out}}}},k \in {{\cal K}}.
\end{align}
Then, we utilize an effective method to transform \eqref{Ar_exp1} into a distributionally robust chance-constraint, given by
\begin{align}\label{Ar_exp2}
	\mathop {\inf }\limits_{{\mathbb{P}} \in {\cal P}} {\Pr}_ {\mathbb{P}}\left\{ {\Delta {{\bm{u}}^T}\Delta {\bm{u}} + 2\Delta {{\bm{u}}^T}\left( {{\bm{\hat u}} - {{\bm{p}}_k}} \right) + {{\left\| {{\bm{\hat u}} - {{\bm{p}}_k}} \right\|}^2}} \right. \nonumber\\
	\left. { - {P_{{\rm{c}},k}}{\gamma _k} \le 0} \right\} \ge 1 - {P_{{\rm{out}}}},k \in {{\cal K}},
\end{align}
where $\mathop {\inf }\limits_{{\mathbb{P}} \in {\cal P}} {\Pr _{\mathbb{P}}}\left\{  \cdot  \right\}$ denotes the lower bound of the probability under the probability distribution ${\mathbb{P}}$ and ${\cal{P}}$ is called ambiguity set, which includes all the possible distributions of the LSE $\Delta {\bm{u}}$. The chance-constraint \eqref{Ar_exp2} satisfies the demand of finding the worst-case distribution among all the possible distributions from the ambiguity set ${\cal{P}}$.

To further deal with this constraint, we introduce a CVaR-based method, which is regarded as a good convex approximation of the worst-case chance-constraint, as shown in the following lemma.
\begin{lemma}{(CVaR-based method)} \cite{Zymler2013MP, Zhang2019JIOT}\label{lemma3}
	Consider a continuous function $g$, which is concave or quadratic in ${\bm{\xi }}$.  The distributionally robust chance-constraint is equivalent to the worst-case constraint, which is given as
	\begin{align}\label{CVaR_lemma1}
		\mathop {\inf }\limits_{{\mathbb{P}} \in {\cal P}} {{\Pr} _{\mathbb{P}}}\left\{ {g\left( {\bm{\xi }} \right) \le 0} \right\} \ge 1 - \varepsilon  \Leftrightarrow \mathop {\sup }\limits_{{\mathbb{P}} \in {\cal P}} {\mathbb{P}} - {\rm{CVa}}{{\rm{R}}_\varepsilon }\left\{ {g\left( {\bm{\xi }} \right)} \right\} \le 0,
	\end{align}
	where ${\mathbb{P}} - {\rm{CVa}}{{\rm{R}}_\varepsilon }\left\{ {g\left( {\bm{\xi }} \right)} \right\}$ denotes the CVaR of ${g\left( {\bm{\xi }} \right)}$ at threshold $\varepsilon $ under distribution ${\mathbb{P}}$, given as
	\begin{align}
		{\mathbb{P}} - {\rm{CVaR}}_{\varepsilon }\left\{ {g\left( {\bm{\xi }} \right)} \right\} = \mathop {\inf }\limits_{\chi  \in {\mathbb{R}}} \left\{ {\chi  + \frac{1}{\varepsilon }{{\mathbb{E}}_{\mathbb{P}}}\left[ {{{\left( {g\left( {\bm{\xi }} \right) - \chi } \right)}^ + }} \right]} \right\},
	\end{align}
	where $\chi $ is an auxiliary variable introduced by CVaR 
	and ${\left( A \right)^ + } \buildrel \Delta \over = \max \left\{ {0,A} \right\}$.
\end{lemma}

Furthermore, by using the following lemma, we can convert the worst-case CVaR on the right hand side of \eqref{CVaR_lemma1} into a group of semidefinite programs (SDPs).
\begin{lemma}\cite{Zymler2013MP, Zhang2019JIOT} \label{lemma4}
	Let $g\left( {\bm{\xi }} \right) = {{\bm{\xi }}^T}{\bm{B\xi }} + {{\bm{b}}^T}{\bm{\xi }} + {{\bm{b}}^0}$ being a quadratic function of ${\bm{\xi}}$, $\forall{\bm{\xi }} \in {{\mathbb{R}}^n}$. The worst-case CVaR can be computed as
	\begin{subequations}
		\begin{align}
		\mathop {\sup }\limits_{{\mathbb{P}} \in {\cal P}} {\mathbb{P}}& - {\rm{CVa}}{{\rm{R}}_\varepsilon }\left\{ {g\left( {\bm{\xi }} \right)} \right\} =  \mathop {\min }\limits_{\chi ,{\bm{M}}} \chi  + \frac{1}{\varepsilon }{\rm{tr}}\left\{ {{\bm{DM}}} \right\}\\
		{\rm{s.t.}}\quad &{\bm{M}} \succeq {\bm{0}},\ {\bm{M}} \in {{\mathbb{S}}^{n + 1}},\\
		&{\chi}  \in {\mathbb{R}},\\
		&{\bm{M}} - \left[ {\begin{array}{*{20}{c}}
			{\bm{B}}&{\frac{1}{2}{\bm{b}}}\\
			{\frac{1}{2}{{\bm{b}}^T}}&{{{\bm{b}}^0} - \chi }
			\end{array}} \right] \succeq {\bm{0}},
		\end{align}
	\end{subequations}
	where $\bm{M}$ is an auxiliary matrix variable, ${{\mathbb{S}}^{n}}$ denotes the space of $n$-dimensional symmetric matrix, and ${\bm{D}}$ is denoted as
	\begin{align}
		{\bm{D}} \buildrel \Delta \over = \left[ {\begin{array}{*{20}{c}}
			{{\bm{\Sigma }} + {\bm{\mu }}{{\bm{\mu }}^T}}&{\bm{\mu }}\\
			{{{\bm{\mu }}^T}}&1
			\end{array}} \right],
	\end{align}
	where ${\bm{\Sigma}} \in {\mathbb{S}}^n$ and ${\bm{\mu}} \in {\mathbb{R}}^n$ are respectively denoted as the covariance matrix and mean vector of random vector ${\bm{\xi}}$.
\end{lemma}

By using the Lemma \ref{lemma3} and Lemma \ref{lemma4}, the constraint \eqref{Ar_exp2} is equivalent to
\begin{subequations}
	\begin{align}
		&\chi  + \frac{1}{P_{\rm{out}} }{\rm{tr}}\left\{ {{\bm{DM}}} \right\} \le 0, \label{Ar_c1}\\
		&{\bm{M}} \succeq {\bm{0}},\ {\bm{M}} \in {{\mathbb{S}}^{4}}, \label{Ar_c2}\\
		&{\chi}  \in {\mathbb{R}}, \label{Ar_c3} \\
		&{\bm{M}} - \left[ {\begin{array}{*{20}{c}}
			{\bm{I}}&{{{\bm{\hat{u}} }- {{\bm{p}}}_k}}\\
			{\left({{{\bm{\hat{u}} }- {{\bm{p}}}_k}}\right)^T}&{{{\left\| {{\bm{\hat u}} - {{\bm{p}}_k}} \right\|}^2} - {P_{{\rm{c}},k}}{\gamma _k} - \chi }
			\end{array}} \right] \succeq {\bm{0}},k \in {{\cal K}},\label{Ar_c4}
	\end{align}
\end{subequations}
where ${\bm{M}}$ and ${\chi}$ are auxiliary variables, and
\begin{align}
	{\bm{D}} \buildrel \Delta \over = \left[ {\begin{array}{*{20}{c}}
		{{\bm{J}}_{\rm{p}}^{ - 1}}&{\bm{0}}\\
		{\bm{0}}^T&1
		\end{array}} \right].
\end{align}

Thus, the robust power allocation problem \eqref{p1Ar} is reformulated as 
\begin{subequations}\label{p1Ar_solve}
	\begin{align}
	\mathop {\min }\limits_{P_{{\rm{s}},k} ,P_{{\rm{c}},k}, {\bm{M}}, {\chi} }&\quad {\rm{tr}}\left\{ {{\bm{J}}_{\rm{p}}^{ - 1}} \right\} \label{p1Ar_solve_1} \\
	{\rm{s}}.{\rm{t}}.&\quad \eqref{p1Ar_5}, \eqref{p1Ar_6}, \eqref{Ar_c1}, \eqref{Ar_c2}, \eqref{Ar_c3}, \eqref{Ar_c4}. \nonumber
	\end{align}
\end{subequations}
However, it is noted that  problem \eqref{p1Ar_solve} is still non-convex due to the fact that the variables $P_{{\rm{s}},k}$ and ${\bm{M}}$ are coupled in the constraint \eqref{Ar_c1}. Similarly, we propose to adopt the AO algorithm to decouple this problem into two subproblems. Specifically, at the $n$th iteration, the two subproblems are optimized as follows:
\subsubsection{Communication Subproblem}
For the given sensing power $P_{{\rm{s}},k}^{\left(n-1\right)}$, the communication power $P_{{\rm{c}},k}^{\left(n\right)}$ and variables  $\left\{{\bm{M}}^{\left(n\right)}, {\chi}^{\left(n\right)}\right\}$ can be updated by solving
\begin{subequations}\label{p1Ar_subp1}
	\begin{align}
	\mathop {\min }\limits_{P_{{\rm{c}},k}, {\bm{M}}, {\chi} }&\quad \sum\limits_{k \in {\cal{K}}} {{P_{{\rm{c}},k}}} \label{p1Ar_subp1_1} \\
	{\rm{s}}.{\rm{t}}.&\quad \eqref{Ar_c1}, \eqref{Ar_c2}, \eqref{Ar_c3}, \eqref{Ar_c4}. \nonumber
	\end{align}
\end{subequations}

\subsubsection{Sensing Subproblem} For the obtained communication power $P_{{\rm{c}},k}^{\left(n\right)}$ and parameters $\left\{{\bm{M}}^{\left(n\right)}, {\chi}^{\left(n\right)}\right\}$, the sensing power $P_{{\rm{s}},k}^{\left(n\right)}$  can be updated by solving
\begin{subequations}\label{p1Ar_subp2}
	\begin{align}
	\mathop {\min }\limits_{P_{{\rm{s}},k}}&\quad {\rm{tr}}\left\{ {{\bm{J}}_{\rm{p}}^{ - 1}} \right\} \label{p1Ar_subp2_1} \\
	{\rm{s}}.{\rm{t}}.&\quad \eqref{p1Ar_5}, \eqref{p1Ar_6}, \eqref{Ar_c1}. \nonumber
	\end{align}
\end{subequations}
However, it is noted that the constraint \eqref{Ar_c1} is likely to be infeasible due to the fact that ${\bm{J}}_{\rm{p}}$ exists  in ${\bm{D}}$ in the form of  an inverse when applying the ToA-based CRB. Hence, we adopt the first-order  Taylor expansion method to approximate the constraint \eqref{Ar_c1} as
\begin{align}\label{pos_subp_const}
	\chi  + \frac{1}{P_{\rm{out}} }\left({\rm{tr}}\left\{ {{\bm{{D}}_0}}{{\bm{M}}} \right\} - {\rm{tr}}\left\{ {{\bm{{D}}_0^2}}\left({\bm{D}}^{-1} - {\bm{D}}_0^{-1}\right){{\bm{M}}} \right\}\right) \le 0,
\end{align}
where ${\bm{D}}_0$ is calculated based on the given initial $P_{{\rm{s}},k}^{\left(0\right)}$. Thus, the sensing subproblem \eqref{p1Ar_subp2} can be reformulated as 
\begin{subequations}\label{p1Ar_subp2-1}
	\begin{align}
	\mathop {\min }\limits_{P_{{\rm{s}},k}}&\quad {\rm{tr}}\left\{ {{\bm{J}}_{\rm{p}}^{ - 1}} \right\} \label{p1Ar_subp2_1-1} \\
	{\rm{s.t.}}&\quad \eqref{p1Ar_5}, \eqref{p1Ar_6}, \eqref{pos_subp_const}. \nonumber
	\end{align}
\end{subequations}

Finally, the above two convex subproblems \eqref{p1Ar_subp1} and \eqref{p1Ar_subp2-1} can be directly solved by CVX . 
The details are listed in Algorithm \ref{alg3}.
\begin{algorithm}[htp]
	\caption{CVaR-AO method for the robust power allocation with the arbitrary distributed LSE}
	\label{alg3}
	\LinesNumbered
	\SetKwRepeat{Repeat}{repeat}{until}
	\SetKw{Ini}{Initialization:}
	\KwIn{$\bar R > 0$, ${P_{\rm{out}}} > 0$, $P_{\rm{total}} > 0$, the tolerance
		of accuracy $\epsilon > 0$, and the maximum iteration number $N_{\rm{max}}$}
	\KwOut{Sensing power $P_{{\rm{s}},k}$ and communciation power $P_{{\rm{c}},k}$}
	\Ini $P_{{\rm{s}},k}^{\left(0\right)} > 0$ and set $n=0$\;
	\Repeat{The decrease of the objective value is below the given tolerance $\epsilon$ {\rm{or}} $n > N_{\rm{max}}$}{
		$n = n + 1$\;
		Update $\left\{P_{{\rm{c}},k}^{\left(n\right)}, {\bm{M}}^{\left(n\right)}, {\chi}^{\left(n\right)}\right\}$ by solving  subproblem \eqref{p1Ar_subp1}\;
		Update $P_{{\rm{s}},k}^{\left(n \right)}$ by solving  subproblem \eqref{p1Ar_subp2-1}\;
	}
	\KwRet $P_{{\rm{s}},k}$ and $P_{{\rm{c}},k}$\;
\end{algorithm}

\subsection{Non-robust ISAC Design}\label{nonrobust_design}
To demonstrate the robustness of  our proposed UAV-aided ISAC system, we also present the non-robust design scheme, which  takes the term ${\bm{\hat h}}_k$ as the perfect CSI without considering the term $\Delta{\bm{h}}_k$. In this case, the power allocation problem of the non-convex ISAC design is formulated as
\begin{subequations}
	\begin{align}
	\mathop {\min }\limits_{P_{{\rm{s}},k} ,P_{{\rm{c}},k} }&\quad {\rm{tr}}\left\{ {{\bm{J}}_{\rm{p}}^{ - 1}} \right\} \\
	{\rm{s.t.}}&\quad {\log _2}\left( {1 + \frac{{{P_{{\rm{c}},k}}{\lambda ^2}\left|{\bm{\hat a}}_k^H{{\bm{w}}_k}\right|^2}}{{16{\pi ^2}{N_{\rm{0}}}{{\left\| {{\bm{\hat u}} - {{\bm{p}}_k}} \right\|}^2}}}} \right) \ge {\bar R},k \in {{\cal K}}, \\
	&\quad \sum\limits_{k \in {{\cal{K}}}} {{P_{{\rm{s}},k}} + {P_{{\rm{c}},k}}} \le P_{\rm{total}},\\
	&\quad {{\bm{J}}}_{\rm{p}} \succeq {\bm{0}}.
	\end{align}
\end{subequations}
The problem is convex that can be directly solved by CVX. 

\begin{appendix}[Derivation of the Formulation   \eqref{location_related_fim}]
	In this work, we adopt the ToA-based method to estimate the location and CRB of the UE. Hence, we simplify the received sensing signal in \eqref{receive_pos_signal} of the $k$th UAV  as
	\begin{align}
	{y_{{\rm{s}},k}}\left( t \right) ={\sqrt{P_{{\rm{s}},k}}}{\alpha _k}{s_{{\rm{s}},k}}\left( {t - {\tau _k}} \right) + {n}\left( t \right),
	\end{align}
	where $\alpha_k \buildrel \Delta \over =  {\bm{h}}_k^H{{\bm{w}}_k}$. 
	
	Let define the unknown parameter set ${\bm{\eta}}$ as 
	\begin{align}
		{\bm{\eta}}  \buildrel \Delta \over = {\left[ {\tau_1, \Re \left\{ {{\alpha _1}} \right\}, \Im \left\{ {{\alpha _1}} \right\}, \cdots ,\tau_K, \Re \left\{ {{\alpha _K}} \right\}, \Im \left\{ {{\alpha _K}} \right\}} \right]^T}.
	\end{align}
	The mean square error (MSE) of unbiased estimation ${{\bm{\hat \eta }}}$ of ${\bm{\eta }}$ is given by
	\begin{align}
	{\mathbb E}\left\{ {\left( {{\bm{\hat \eta }} - {\bm{\eta }}} \right){{\left( {{\bm{\hat \eta }} - {\bm{\eta }}} \right)}^{\rm{T}}}} \right\} \succeq {\bm{J}}_{\bm{\eta }}^{ - 1},
	\end{align}
	where ${{\bm{J}}_{\bm{\eta}} }$ is the channel-related FIM. For ease descriptions, we merge the received sensing signals from multiple UAVs into a vector, given as ${{\bm{y}}_{\rm{s}}}\left( t \right) \buildrel \Delta \over = {\left[ {{y_{{\rm{s}},1}}\left( t \right), \cdots ,{y_{{\rm{s}},K}}\left( t \right)} \right]^T}$, Then, the channel-related FIM ${{\bm{J}}_{\bm{\eta}} }$ is given by
	\begin{align}\label{fim_element}
	{\left[ {{{\bm{J}}_{\bm{\eta}} }} \right]_{l,m}} = \frac{2}{{{N_0}}}\int_0^{{T_{\rm{s}}}} {\Re \left\{ {\frac{{\partial {{\bm{m }}^H}\left( t \right)}}{{\partial l}}\frac{{\partial {\bm{m }}\left( t \right)}}{{\partial m}}} \right\}dt},
	\end{align}
	where $l,m \in {\bm{\eta}} $, and
	\begin{align}
	{\bm{m }}\left( t \right) \buildrel \Delta \over = \left[ {\begin{array}{*{20}{c}}
		{{\alpha _1}{s_{{\rm{s}},1}}\left( {t - {\tau _1}} \right)}\\
		\vdots \\
		{{\alpha _K}{s_{{\rm{s}},K}}\left( {t - {\tau _K}} \right)}
		\end{array}} \right].
	\end{align}
	
	After taking the partial derivation of \eqref{fim_element}, we can express ${{\bm{J}}_{\bm{\eta}} }$ as
	\begin{align}
	{{\bm{J}}_{\bm{\eta }}} \buildrel \Delta \over = {\rm{diag}}\left\{ {{{\bm{\Phi }}_1}, \cdots ,{{\bm{\Phi }}_K}} \right\},
	\end{align}
	where
	\begin{align}\label{eq33}
	{{\bm{\Phi }}_k} &= \left[ {\begin{array}{*{20}{c}}
		{\frac{{8{\pi ^2}{P_{{\rm{s}},k}}{{\left| {{\alpha _k}} \right|}^2}{\beta} ^2}} {{{N_0}}}}&{{{\bm{0}}}}\\
		{{\bm{0}}^T}&{\frac{2}{{{N_0}}}{{\bm{I}}}}
		\end{array}} \right],k \in {\cal K}.
	\end{align}
	In \eqref{eq33}, $\beta$ denotes the effective bandwidth, which is given by
	\begin{small}
		\begin{align}
		{\beta}^2 \buildrel \Delta \over = \int_{ - \infty }^{ + \infty } {{{\left| {fS\left( f \right)} \right|}^2}df}  = \int_{ 0 }^{ T_{\rm{s}} } {\frac{{\partial {s_{{\rm{s}},k}}\left( {t - {\tau _k}} \right)}}{{\partial {\tau _k}}}\frac{{\partial {s_{{\rm{s}},k}}\left( {t - {\tau _k}} \right)}}{{\partial {\tau _k}}}dt},
		\end{align}
	\end{small}
	where ${S\left( f \right)} $ denotes as the Fourier transform of ${s_{{\rm{s}},k}}\left( t \right)$.
	
	Moreover, by using the transform matrix ${\bm{\Upsilon}}  = \frac{{\partial {\bm{\eta }}}}{{\partial {\bm{\tilde \eta }}}}$, we transform the channel-related FIM ${{\bm{J}}_{\bm{\eta }}}$ to the location parameter involved  FIM  ${\bm{J}}_{\bm{\tilde \eta }} $, where ${\bm{\tilde \eta }}$ is defined as the  location-related unknown parameter vector, given by
	\begin{align}
		{\bm{\tilde \eta }} \buildrel \Delta \over = {\left[ {{\bm{u}}^T},\Re \left\{ {{\alpha _1}} \right\}, \Im \left\{ {{\alpha _1}} \right\}, \cdots , \Re \left\{ {{\alpha _K}} \right\}, \Im \left\{ {{\alpha _K}} \right\} \right]^T}.
	\end{align} 
	In this case, the location parameter involved  ${\bm{J}}_{\bm{\tilde \eta }} $ can be expressed as
	\begin{align}
	{{\bm{J}}_{\bm{\tilde \eta }}} = {\bm{\Upsilon}} {{\bm{J}}_{{\bm{ \eta }}}}{{\bm{\Upsilon}} ^{\rm{T}}},
	\end{align}
	where
	\begin{subequations}
		\begin{align}
		{\bm{\Upsilon}}  &= \left[ {\begin{array}{*{20}{c}}
			{{{\bm{U}}_1}}& \cdots &{{{\bm{U}}_K}}\\
			{{{\bm{T}}_1}}&{}&{}\\
			{}& \ddots &{}\\
			{}&{}&{{{\bm{T}}_K}}
			\end{array}} \right],\\
		{{\bm{U}}_k} &= \left[ {\frac{{\partial {\tau _k}}}{{\partial {\bm{u}}}},{\bm{0}}} \right],\\
		{{\bm{T}}_k} &= \left[ {\begin{array}{*{20}{c}}
			0&1&0\\
			0&0&1
			\end{array}} \right].
		\end{align}
	\end{subequations}
	Then, we can express the ${\bm{J}}_{\bm{\tilde \eta }} $ as
	\begin{align}
	{{\bm{J}}_{{\bm{\tilde \eta }}}} = \left[ {\begin{array}{*{20}{c}}
		\sum\limits_{k \in {\cal K}} {{{\bm{U}}_k}{{\bm{\Phi }}_k}{\bm{U}}_k^T}&{\bm{0}}\\
		{{{\bm{0}}^T}}&{\frac{2}{N_0}\bm{I}}
		\end{array}} \right].
	\end{align}
	
	According to the Schur complement, the location-related FIM ${\bm{J}}_{\rm{p}}$ is given by
	\begin{align}
	{\bm{J}}_{\rm{p}} = \sum\limits_{k \in {\cal K}} {{\frac{{8{\pi ^2}{P_{{\rm{s}},k}}{{\left| {{\alpha _k}} \right|}^2}{\beta} ^2}} {{{N_0}}}}{\frac{{\partial {\tau _k}}}{{\partial {\bm{u}}}}\frac{{\partial {\tau _k}}}{{\partial {\bm{u}}^T}}}}.
	\end{align}
	Finally, the formulation of \eqref{location_related_fim} can be obtained.
	
\end{appendix}

\bibliographystyle{IEEE-unsorted}

\bibliography{refsJCPS}

\end{document}